\documentclass[tradiabstract]{aa} 
\usepackage{graphicx}
\usepackage{natbib}
\usepackage{subfigure}
\bibpunct{(}{)}{;}{a}{}{,}
\begin{document}
   \title{Relating basic properties of bright early-type dwarf galaxies
   to their location in Abell 901/902}

   \author{Fabio D. Barazza\inst{1} \and Christian Wolf\inst{2} \and
     Meghan E. Gray\inst{3} \and Shardha Jogee\inst{4} \and Michael
     Balogh\inst{5} \and Daniel H. McIntosh\inst{6} \and David
     Bacon\inst{7} \and Marco Barden\inst{8} \and Eric F. Bell\inst{9}
     \and Asmus B\"ohm\inst{8,15} \and John A.R. Caldwell\inst{10} \and
     Boris H\"aussler\inst{3} \and Amanda Heiderman\inst{4} \and
     Catherine Heymans\inst{11} \and Knud Jahnke\inst{9} \and Eelco
     van Kampen\inst{12} \and Kyle Lane\inst{3} \and Irina
     Marinova\inst{4} \and Klaus Meisenheimer\inst{8} \and Chien Y.
     Peng\inst{13} \and Sebastian F. Sanchez\inst{14} \and Andy
     Taylor\inst{10} \and Lutz Wisotzki\inst{15} \and
     Xianzhong Zheng\inst{16}
          }

   \institute{Laboratoire d'Astrophysique, \'Ecole
Polytechnique F\'ed\'erale de Lausanne (EPFL), Observatoire de
Sauverny CH-1290 Versoix, Switzerland \email{fabio.barazza@epfl.ch}
\and
Department of Physics, Denys Wilkinson Building, University of
Oxford, Keble Road, Oxford, OX1 3RH, UK
\and
School of Physics and Astronomy, The University of Nottingham,
University Park, Nottingham NG7 2RD, UK
\and
Department of Astronomy, University of Texas at Austin, 1 University
Station, C1400 Austin, TX 78712-0259, USA
\and
Department of Physics and Astronomy, University Of Waterloo, Waterloo,
Ontario, N2L 3G1, Canada
\and
Department of Physics, 5110 Rockhill Road, University of
Missouri--Kansas City, Kansas City, MO 64110, USA
\and
Institute of Cosmology and Gravitation, Dennis Sciama Building,
Burnaby Road, Portsmouth, PO1 3FX, UK
\and
Institute for Astro- and Particle Physics, University of Innsbruck,
Technikerstr. 25/8, A-6020 Innsbruck, Austria
\and
Max-Planck-Institut f\"{u}r Astronomie, K\"{o}nigstuhl 17, D-69117,
Heidelberg, Germany
\and
University of Texas, McDonald Observatory, Fort Davis, TX 79734, USA
\and
SUPA, Institute for Astronomy, University of Edinburgh, Royal
Observatory, Edinburgh EH9 3HJ, UK
\and
ESO, Karl-Schwarzschild-Str. 2, D-85748, Garching bei München, Germany
\and
NRC Herzberg Institute of Astrophysics, 5071 West Saanich Road,
Victoria V9E 2E7, Canada
\and
Centro Hispano Aleman de Calar Alto, C/Jesus Durban Remon 2-2, E-04004
Almeria, Spain
\and
Astrophysikalisches Insitut Potsdam, An der Sternwarte 16, D-14482
Potsdam, Germany
\and
Purple Mountain Observatory, National Astronomical Observatories,
Chinese Academy of Sciences, Nanjing 210008, PR China
}

   \date{Received; accepted}

 
  \abstract
   {We present a study of the population of bright early-type dwarf
     galaxies in the multiple-cluster system Abell 901/902. We use data
     from the STAGES survey and COMBO--17 to investigate the relation
     between the color and structural properties of the dwarfs and
     their location in the cluster. The definition of the dwarf sample
     is based on the central surface brightness and includes galaxies
     in the luminosity range $-16\ge M_B \ga-19$ mag. Using a fit to
     the color magnitude relation of the dwarfs, our sample is divided
     into a red and blue subsample. We find a color-density relation
     in the projected radial distribution of the dwarf sample: at the
     same luminosity dwarfs with redder colors are located closer to
     the cluster centers than their bluer counterparts. Furthermore,
     the redder dwarfs are on average more compact and rounder than
     the bluer dwarfs. These findings are consistent with theoretical
     expectations assuming that bright early-type dwarfs are the
     remnants of transformed late-type disk galaxies involving
     processes such as ram pressure stripping and galaxy
     harassment. This indicates that a considerable fraction of
     dwarf elliptical galaxies in clusters are the results of
     transformation processes related to interactions with their host
     cluster.

   \keywords{}}
   \titlerunning{Early-type dwarfs in Abell 901/902}
   \maketitle
%

\section{Introduction}
Early-type dwarf galaxies appear to be spheroidal stellar systems with
no signs of ongoing star formation, dust content, or substructures,
except a compact high surface brightness nucleus in the center, which
is found in a large fraction of early-type dwarfs
\citep{cal83,bin91,ryd99,bar03,lis06}. Whether early-type dwarfs
constitute a genuine class of galaxies or are merely the counterparts
of giant elliptical galaxies at lower luminosities is a matter of
ongoing debate. The interpretation of early-type dwarfs as smaller and
fainter versions of giant ellipticals is based on the continuity of
some structural parameters \citep{jer97,gra03a,tru04,gav05,fer06},
while the dichotomy picture rests upon the assumption that early-type
giants and dwarfs have different formation origins \citep[merger
  vs. transformation;][]{wir84,kor85,kor09}. In the latter case
early-type dwarfs are often referred to as spheroidals, in order to
distinguish them from ellipticals. We refer to our sample of
early-type dwarfs as dwarf ellipticals (dEs), without giving
preference to either of the two interpretations.

One of the most important aspects of dEs is that they are almost
exclusively located in regions of high galaxy density. In fact, they
can be regarded as the typical galaxy type in galaxy clusters, where
they numerically dominate the total galaxy population
\citep{bin85,fer89,fer94}. On the other hand they seem to be almost
absent in low density regions outside of clusters and galaxy groups
\citep{bin90,mat98,hai07}. Therefore, the formation and evolution of
dEs seems to be intimately connected to their environment. The most
likely processes believed to be dominant in high density environments
are ram pressure stripping \citep{gun72,lin83,aba99,zee04}, tidal
shaking or stirring \citep{may01b}, and galaxy harassment
\citep{moo98,mas05}. These processes can strongly affect infalling
galaxies causing substantial mass loss, the termination of star
formation, probably preceded by a final starburst, and transform the
galaxies morphologically \citep[e.g.,][]{bos06}. If the structure and
photometric properties of dEs are due to such transformation
processes, the progenitors might be significantly different. On
the other hand, there also remains the possibility that a certain
fraction of dEs belong to the primordial cluster galaxy population
and never suffered from transformations.

In order to connect the formation of dEs to their environment,
observational indicators have to be found, which are very likely to be
caused by interactions with the environment. One promising route is to
identify properties which the dEs may have inherited from their
progenitors, and which have not been erased completely during the
transformation. The hidden disk, spiral, and bar features discovered
in a number of dEs in different clusters are believed to be such
remnant structures \citep{jer00,bar02,gra03b,der03,lis06}. These
structures are only visible, when image processing methods such as
unsharp masking or model subtractions are applied, and indicate that
the dEs are the products of transformations caused by the interactions
of disk galaxies with the cluster and its members. Supporting this
picture is the observation that some dEs themselves
\citep{der01,ped02,geh03,zee04} or their globular cluster systems are
rotating \citep{bea06,bea09}, which indicates their origin as disky
systems.

Another indication connecting the origin of dEs to their environment
are relations between properties of the dE population and measures of
the local galaxy density. Such relations are expected according to
simulations of transformation processes in a cluster environment
\citep{moo98,mas05,ton07}. Observational indications for corresponding
properties of dE populations in specific galaxy clusters have been
found in a number of studies, using different approaches. \cite{sec96}
found a significant color gradient in the projected radial
distribution of a sample of $\sim250$ dEs in the Coma cluster. This
color gradient was interpreted as a metallicity gradient caused by the
confinement pressure of the intracluster gas \citep{bab92}. A similar
result was reported by \cite{rak04}, who found nucleated dEs (dENs) to
be redder than non-nucleated dEs. The former are known to exhibit a
stronger clustercentric concentration than the latter
\citep{van86,bin87} resulting again in a color gradient. Based on the
different color properties of dENs they were also found to be on
average 5 Gyr older than dEs \citep{rak04}. A different approach
was chosen by \cite{mic08} and \cite{smi09}, who used spectroscopic
observations of dEs in different clusters to find an age gradient in
the sense that older dEs are located closer to the cluster centers
than their younger counterparts. Such gradients alone do not
prove an environmental transformation, but are predicted by models of
galaxy interactions in clusters and can therefore be compared
quantitatively with these predictions.

In general, the color or age gradients found are rather shallow and
the scatters of the individual points are large. We try to improve on
this situation with the present study. We use HST/ACS images from the
Space Telescope A901/902 Galaxy Evolution Survey
\citep[STAGES,][]{gra09} and spectral energy distribution (SED) fits
from COMBO--17 \citep{wol03} to investigate the color and structural
properties of the dE population and their relation to the cluster
environment. Our sample of 470 early-type cluster galaxies also
includes giant ellipticals, which we use as a comparison sample. The
analysis is based on color, central surface brightness, S\'ersic
index, and concentration index. In Sect. \ref{samsel} we describe the
data used and the definition of the sample. In Sect. \ref{colmag},
\ref{dist}, and \ref{compa} we discuss the color-magnitude relation of
the sample, the distribution of the galaxies in the cluster region,
and the compactness of the objects, respectively. A discussion follows
in Sect. \ref{discu} and a summary in Sect. \ref{sum}. Throughout the
paper, we assume a flat cosmology with
$\Omega_M=1-\Omega_{\Lambda}=0.3$ and $H_0=70$ km~s$^{-1}$ Mpc$^{-1}$.

\section{Sample selection from STAGES}\label{samsel}
\subsection{The data from the STAGES survey}
STAGES is a multiwavelength survey aimed at the study of galaxy
formation and evolution in different environments. It covers an area
of $0.5^{\circ}\times 0.5^{\circ}$ centered on the multiple-cluster
system Abell 901(a,b)/902. The multiple-cluster system A901/902 at
$z=0.165$ consists of four galaxy accumulations of different sizes:
three clusters and one group \citep{gra02}. The principal dataset used
in our study is the 80--orbit HST ACS $F606W$ mosaic, which provides
images with a scale of $0\farcs03$/pixel and a point spread function
(PSF) of $0\farcs1$, corresponding to $\sim282$ pc at $z=0.165$. For all
galaxies photometric redshifts are available from COMBO--17
\citep{wol03,wol05} with $\delta z/(1+z)\sim0.007-0.022$ for our
sample \citep{gra09}. Supplementary to the optical data dark matter
maps from weak lensing \citep{hey08}, total star formation rates based
on UV and $Spitzer$ observations \citep{bel05}, and stellar masses
\citep{bor06} are available for the cluster field.
\begin{figure*}[t]
\centering
\subfigure
{
    \includegraphics[width=9cm]{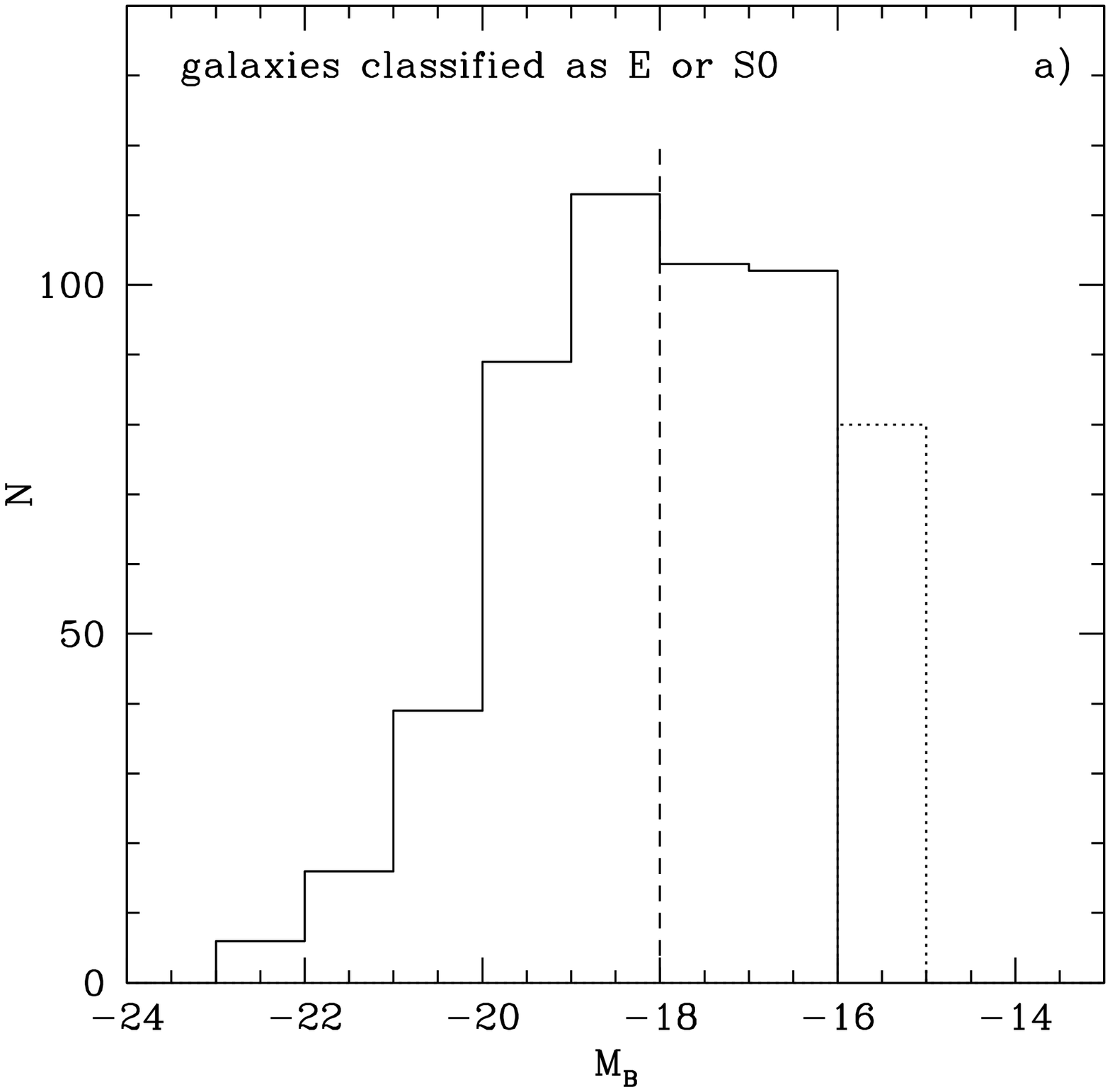}
}
\hspace{-0.5cm}
\subfigure
{
    \includegraphics[width=9cm]{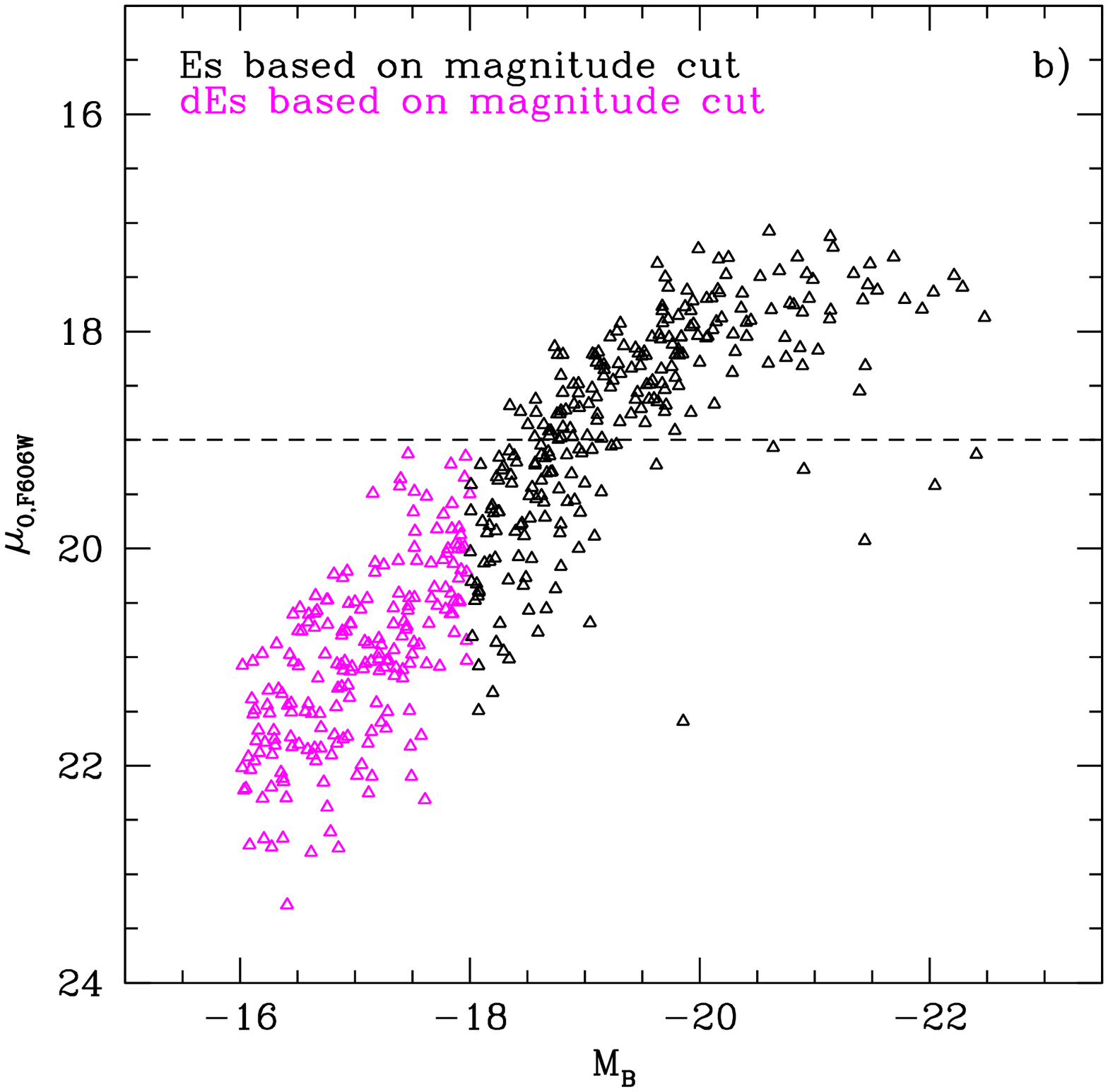}
}
\caption{{\it a)} The $M_B$ distribution of all early-type
  galaxies (solid+dotted histogram). The solid histogram shows the
  sample considered in our study. The dotted line indicates a possible
  separation between giants and dwarfs. {\it b)} Absolute $B-$band
  magnitude versus central surface brightness measured on the
  $F606W-$band images within the central $\sim420$ pc of the
  galaxies. The dashed line indicates a possible separation between
  giants and dwarfs.}
\label{samcut}
\end{figure*}

In addition to the multiwavelength dataset described above, the STAGES
imaging data were processed with the data pipeline GALAPAGOS \citep[Galaxy
Analysis over Large Areas: Parameter Assessment by GALFITting Objects
from SExtractor;][]{bar09}. Part of this pipeline is an analysis of
the surface brightness profiles of the galaxies based on GALFIT
\citep{pen02}, which provides structural parameters such as effective
radii ($r_e$) and S\'ersic indices ($n$) for all galaxies. The latter
is used in our study as a measure of the compactness of our
galaxies. In addition, we also use the CAS concentration index $C$
\citep{ber00} derived using the CAS code \citep{con00,con03} by
\cite{hei09}. See \cite{gra09} for a detailed description of the
entire data reduction and processing work.

We are only interested in cluster galaxies and our initial sample is
based on the general definition of the cluster sample applied to the
STAGES catalog \citep{gra09}. The membership criterion is based on the
photometric redshifts. Galaxies with $z_{\rm cluster}-\Delta z <
z_{\rm galaxy} < z_{\rm cluster}+\Delta z$ are considered to be cluster
members, where $\Delta z$ depends on the total apparent $R$ magnitude
and is defined as
\begin{displaymath}
\Delta z(R) = \sqrt{0.015^2+0.0096525^2(1+10^{0.6(R_{\rm TOT}-20.5)})}.
\end{displaymath}
This photo-z width assures a completeness of $>90\%$ down to
$R_{\rm Vega}=24$ mag \citep[for details see][]{gra09}. This definition
results in a cluster sample of 1990 galaxies. Our final definition of
the early-type cluster members and the subsample of dwarf galaxies is
described below.

\subsection{Visual classifications}
All galaxies in the final STAGES catalog with $z<0.4$ and $R<23.5$ mag
have been visually classified (5090 objects in total). The
classification was performed by seven co-authors (AA-S, FDB, MEG, KJ,
KL, DHM, CW) making sure that each galaxy was inspected by at least
three classifiers. Thus, not all galaxies have been classified by
all classifiers, but for each galaxy at least three independent
classifications are available. Each galaxy was assigned a Hubble type,
including the barred subclasses, representing an average of all available
classifications. The agreement among the classifiers was in general
very good and we base our selection of the parent sample of early-type
galaxies on this classification.

\subsection{The definition of the dE sample}\label{defsam}
Based on the cluster sample of 1990 galaxies and their visual
classification we define our sample of dEs. First, we select all
galaxies classified as E or S0. Since the distinction between Es and
S0s is rather difficult, particularly for the fainter dwarf early-type
galaxies, we match the E and S0 classes and obtain a sample of 550
early-type galaxies. We note that visual types are only available for
1713 galaxies, since only objects with $R<23.5$ mag have been visually
classified. This concerns galaxies with $M_B\ga-15.5$ mag, which
would be excluded anyway (see below). As shown in \citet[Fig. 14]{gra09}
the contamination by field galaxies increases sharply for
objects with $R>22$ mag. We therefore apply a lower magnitude cut at
$M_B\le-16$ mag, where the contamination is estimated to be
$\sim35\%$. This reduces the sample to 470 objects, which constitute
our basic sample considered in the analysis below.
\begin{figure*}
\centering
\subfigure
{
    \includegraphics[width=9cm]{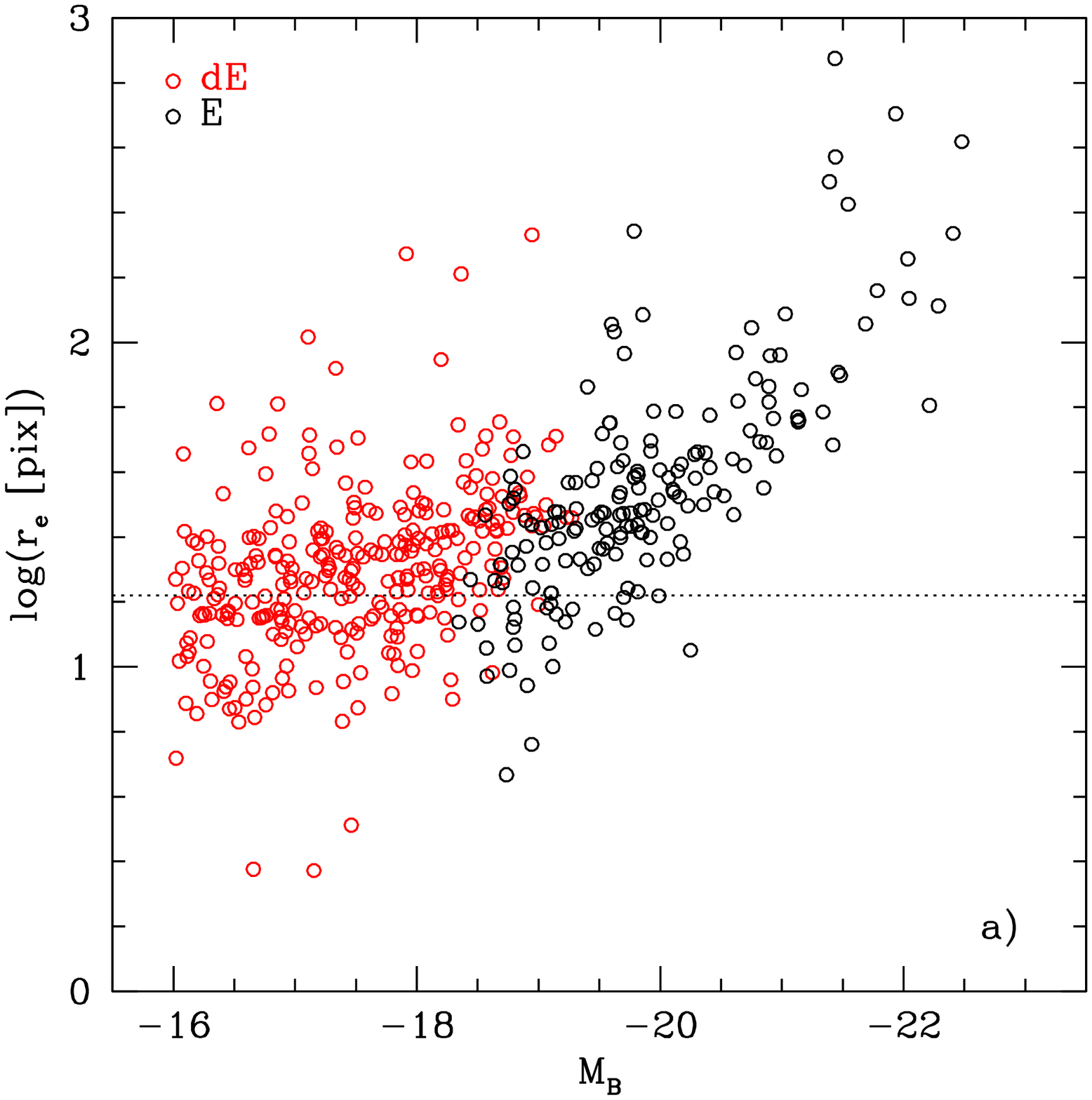}
}
\hspace{-0.5cm}
\subfigure
{
    \includegraphics[width=9cm]{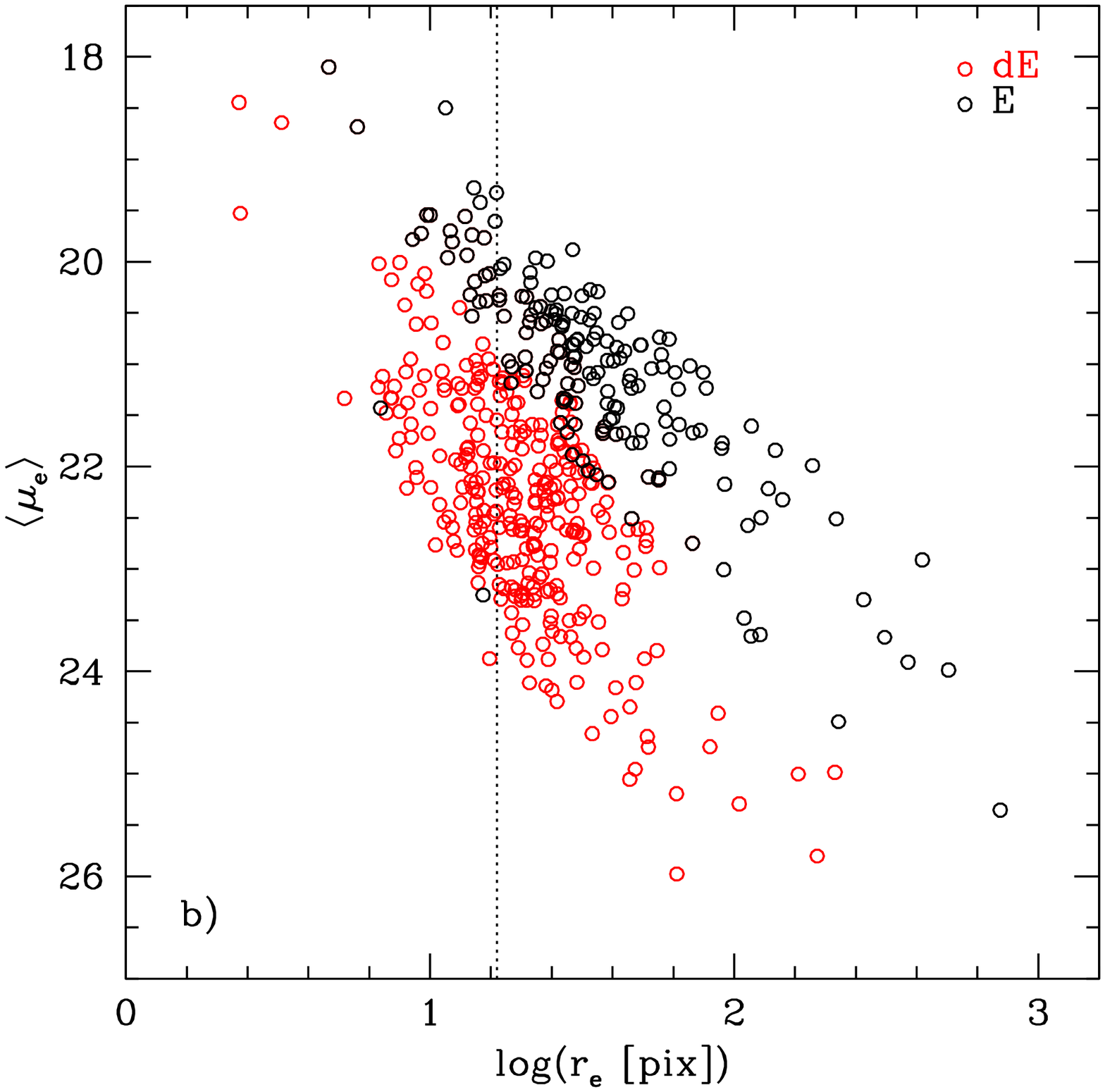}
}
\caption{Two additional photometric scaling relations for the
  early-type dwarf sample ($\mu_{\rm 0,F606W}>19$ mag) and the giant
  early-type sample $\mu_{\rm 0,F606W}\leq19$ mag. {\it a)} The absolute
  $B-$band magnitude versus the effective radius ($r_e$) determined
  by a S\'ersic fit. The dotted line indicates $r_e=1.41$ kpc. {\it
    b)} $r_e$ versus the mean surface brightness
  ($\langle\mu_e\rangle$) within $r_e$. The dotted line indicates
  $r_e=1.41$ kpc.}
\label{korms}
\end{figure*}

Our basic sample includes all cluster galaxies with $M_B\le-16$ mag
classified as either E or S0, which we now want to separate into giant
E and dwarf E\footnote{In the remainder of the paper we only use the
  term E or dE subsuming S0s and dS0s}. The original method to
distinguish dEs from Es is based on morphological grounds using the
fact that dEs have a relatively low central surface brightness and a
rather flat surface brightness profile compared to giant Es
\citep{san84,fer90,fer94}. On images with sufficient resolution and
depth these features can be visually gauged and the classification can
therefore be done by eye. As a result the transition from giants to
dwarfs occurs around an absolute magnitude in the $B$-band of
$M_B\sim-18$ mag \citep{san85}, with some overlap of the two
classes. Therefore, absolute magnitudes in the range $-17<M_B<-19$ mag
are often used to discriminate between dEs and Es \citep{agu05,mat05}
or even between dwarf and giant galaxies in general
\citep{sch90,pat96,zee01,bar06,bos08a}.

Our data do not allow to accurately separate dEs from Es visually and
we have therefore to rely on another measure. As mentioned above, it
is clear that applying a magnitude cut at $M_B=-18$ mag will not
cleanly distinguish between dwarfs and giants: some giants will be in
the dwarf sample and some dwarfs will be missing. This was also shown
in a detailed study of the surface brightness profiles of early-type
galaxies in the Virgo cluster by \cite{kor09}. Dwarf and giant
subsamples exhibit some overlap in absolute magnitude, but separate
quite clearly with respect to central surface brightness \citep[see
  Fig. 34 in][]{kor09}. This suggests that the central surface
brightness is a better measure to define a dwarf sample. Figure
\ref{samcut} illustrates sample selections based on absolute magnitude
and central surface brightness. Figure \ref{samcut}a shows the $M_B$
distribution of all early-type galaxies in A901/902 (solid+dotted
histogram). The solid histogram indicates the sample considered in
our study. A possible sample selection is indicated by the dashed
line. Figure \ref{samcut}b shows $M_B$ versus the central surface
brightness, $\mu_{\rm 0,F606W}$. The latter has been determined using the
central 25 pixels, corresponding to a diameter of $\sim420$ pc. The
magenta triangles show the dwarf sample as defined in Figure
\ref{samcut}a. This definition would exclude a number of galaxies with
rather low $\mu_{\rm 0,F606W}$. A cut applied at $\mu_{\rm 0,F606W}=19$
mag/arcsec$^2$ includes all galaxies with $M_B>-18$ mag and leads to two
distinct sequences in a plot of effective radius versus effective surface
brightness (see below, Fig. \ref{korms}b). It also provides a division
with smaller overlap between the samples, i.e. with less contamination, and
is therefore a cleaner definition of the dwarf sample. Hence, we regard all
early-type galaxies having $\mu_{\rm 0,F606W}>19$ mag/arcsec$^2$ as dEs,
except seven galaxies with $\mu_{\rm 0,F606W}>19$ mag/arcsec$^2$ and
$M_B\leq-19.5$ mag, which we regard as being giant ellipticals with
exceptionally low central surface brightnesses. Our sample of dEs therefore
includes 295 objects.

For the sake of completeness we show in Fig. \ref{korms} two
additional photometric scaling relations. In Fig. \ref{korms}a we plot
the absolute $B$--band magnitude versus the effective radius. The
dashed line indicates an effective radius of $r_e=1.41$ kpc, which was
found to be the typical size of dEs in the Virgo cluster
\citep{bos08b}. Slightly smaller sizes were found for dE populations
in the Antlia \citep{sca08} and Coma \citep{gra03a} clusters, however
with significantly smaller samples sizes. Figure \ref{korms}a shows
that among Es brighter galaxies have large effective radii. For dEs
the scatter is much larger and a similar sequence cannot be clearly
identified. This is primarily caused by the rather small magnitude
range covered by our dE sample \citep[for a comparison see,
  e.g.,][Fig. 2]{der09}. In Fig. \ref{korms}b we plot the effective
radius versus the mean surface brightness within $r_e$. Giants and
dwarfs form two separated sequences of decreasing surface brightness
with increasing effective radius. The sequence for Es is a possible
projection of the fundamental plane of elliptical galaxies and the
fact that we obtain two distinct sequences in this plot supports our
definition of dwarfs and giants. However, the dE sequence is also
caused by the fact that we exclude galaxies with $M_B>-16$ mag,
which would otherwise appear in the lower left corner of the plot
\begin{figure}
\centering
\includegraphics[width=9cm]{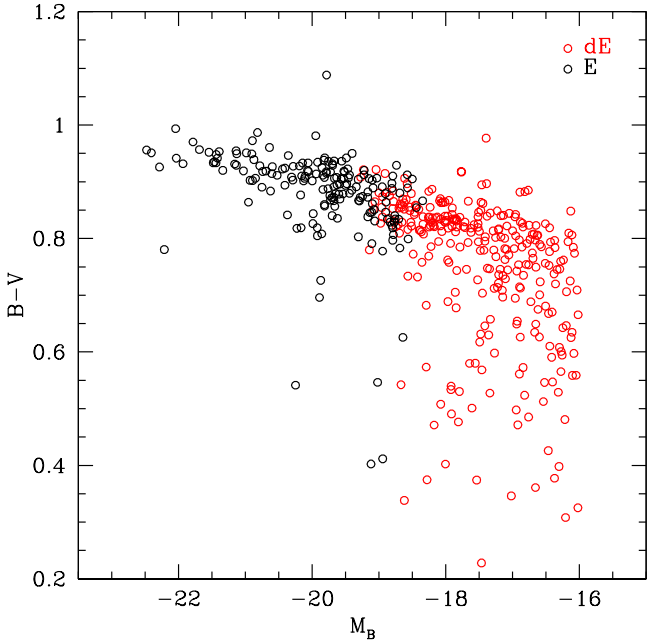}
\caption{The color-magnitude diagram for our sample of early-type
  galaxies. The dwarf and giant subsamples are defined as discussed in
Sect. \ref{defsam}. The typical error of the $B-V$ color is in the
range $0.10-0.15$ mag.}
\label{cmd}
\end{figure}

\section{The color-magnitude relation}\label{colmag}
\begin{figure*}
\centering
\includegraphics[width=15.2cm]{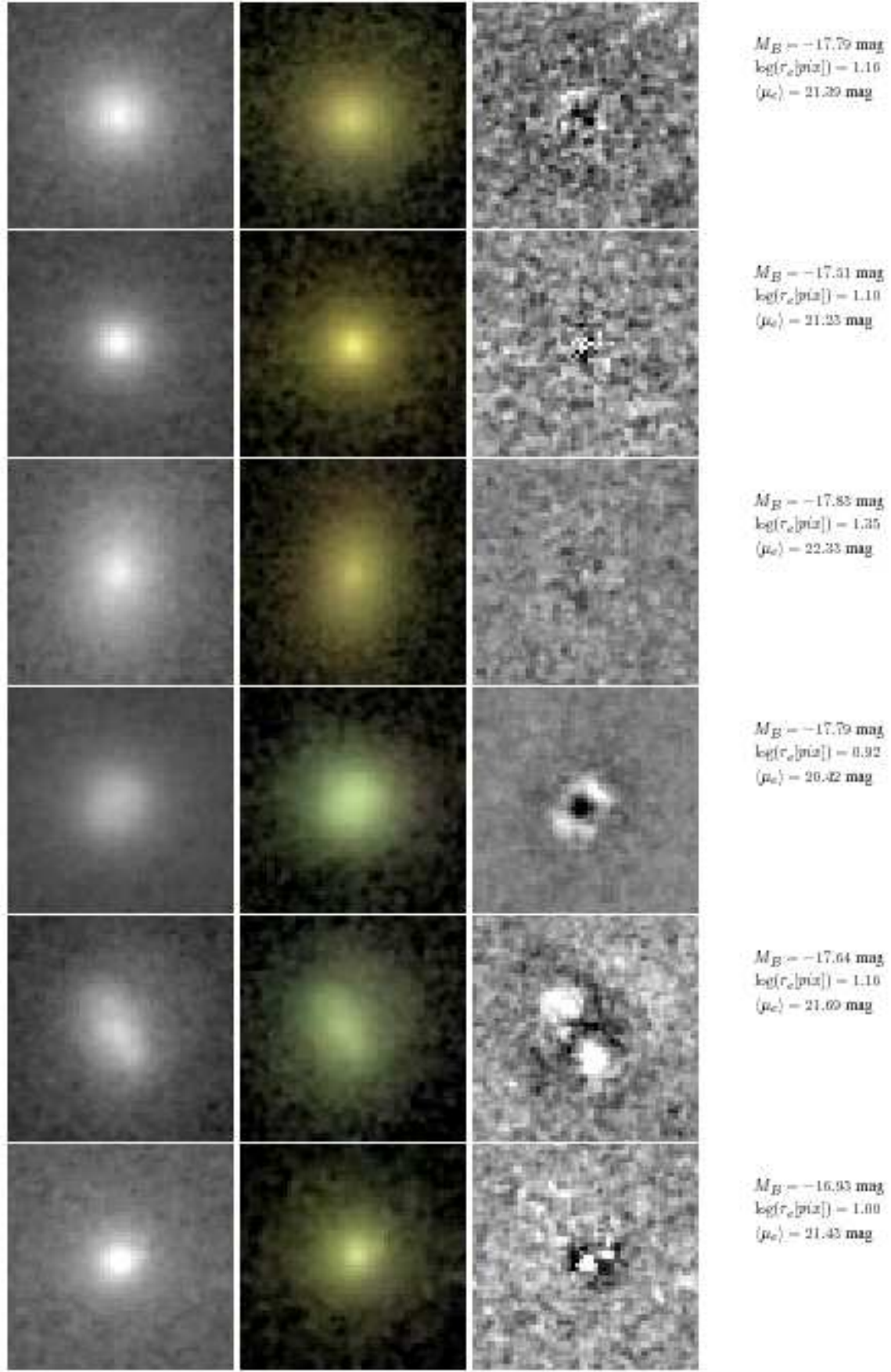}
\caption{Examples of dEs on the red sequence and dEs with large
  separations from the red sequence. First column: the $F606W-$band
  stamp; second column: a color representation of the HST images based
  on $BVR$ colors from COMBO--17 of the same galaxies; third column:
  residual after subtracting a S\'ersic model of the same
  galaxies; fourth column: photometric parameters as plotted in Fig.
  \ref{korms}. The top three rows show dEs on the red sequence and the
  bottom three rows show dEs with large separations from the red
  sequence. All images are $\sim4.2$ kpc on a side.}
\label{ima}
\end{figure*}

Early-type galaxies are typically forming the red sequence in a plot
of absolute magnitude versus color (the color-magnitude diagram,
CMD). In Fig. \ref{cmd} we show such a plot for our early-type galaxy
sample. In fact, Es and a majority of dEs form a rather tight relation
in this plot. However, there are a number of dEs, which are
significantly bluer than expected based on their magnitude. We
emphasize that these objects have been classified as early-type
galaxies, thus they appear to be blue dEs. They are therefore different
than the dEs with blue {\it central} colors observed in local dwarf
galaxy samples \citep{lis06,tul08}. We have reinspected all galaxies, with
large separations from the red sequence and found that their morphological
classification is in general correct. This is illustrated in Fig. \ref{ima}. The top
three rows show dEs on the red sequence. Each galaxy is represented by
a $F606W$--band image (first column), a color representation of the HST
images based on $BVR$ colors from COMBO--17 (second column), and by
the residual after subtracting a S\'ersic model from the $F606W$--band
image (third column). The bottom three rows show the same for dEs
with significantly bluer colors than expected based on their absolute magnitude.
The visual classification was performed using the $F606W$--band images.
In particular the $BVR$--based representations demonstrate the different colors.

A possible reason for the blue colors of these objects could be a
wrong redshift. The contamination of our dE sample by field galaxies
is on the order of $<10\%$ down to $M_B\sim -19$ mag and $\sim35\%$ at
$M_B=-16$ mag \citep{gra09}. The possible contamination therefore roughly
corresponds to the fraction of galaxies below the red sequence. For
instance, a disk galaxy effectively located at a higher redshift could
appear very similar to a closer dE, while maintaining its blue
color. This interpretation is supported by the residual images shown
in the third column of Fig. \ref{ima}. These residuals were obtained
by subtracting the best fitting S\'ersic model from the $F606W$--band
image. While the red dEs hardly show any residual (top three rows), the
blue dEs exhibit strong residuals resembling rings, bars, or even
spiral arms (bottom three rows). We emphasize that such strong
residuals are found in a large majority of blue dEs.

These rather blue dEs with large separations from the red sequence
are either very interesting objects in terms of possible formation
scenarios of dEs or they are simply contaminants, which effectively
are at higher redshifts. We cannot distinguish between these two
possibilities on an individual basis, but we assume that a majority
are not cluster members. We will obtain a better assessment of their
cluster membership through the analysis of low resolution prism
spectroscopy with Magellan (IMACS/PRIMUS) for galaxies down to
$R=23.6$ mag. These observations have already been performed and the
data analysis is underway. For now we include these galaxies in the
following analysis and show them as a separate subsample as defined
below.

\begin{figure}
\centering
\includegraphics[width=9cm]{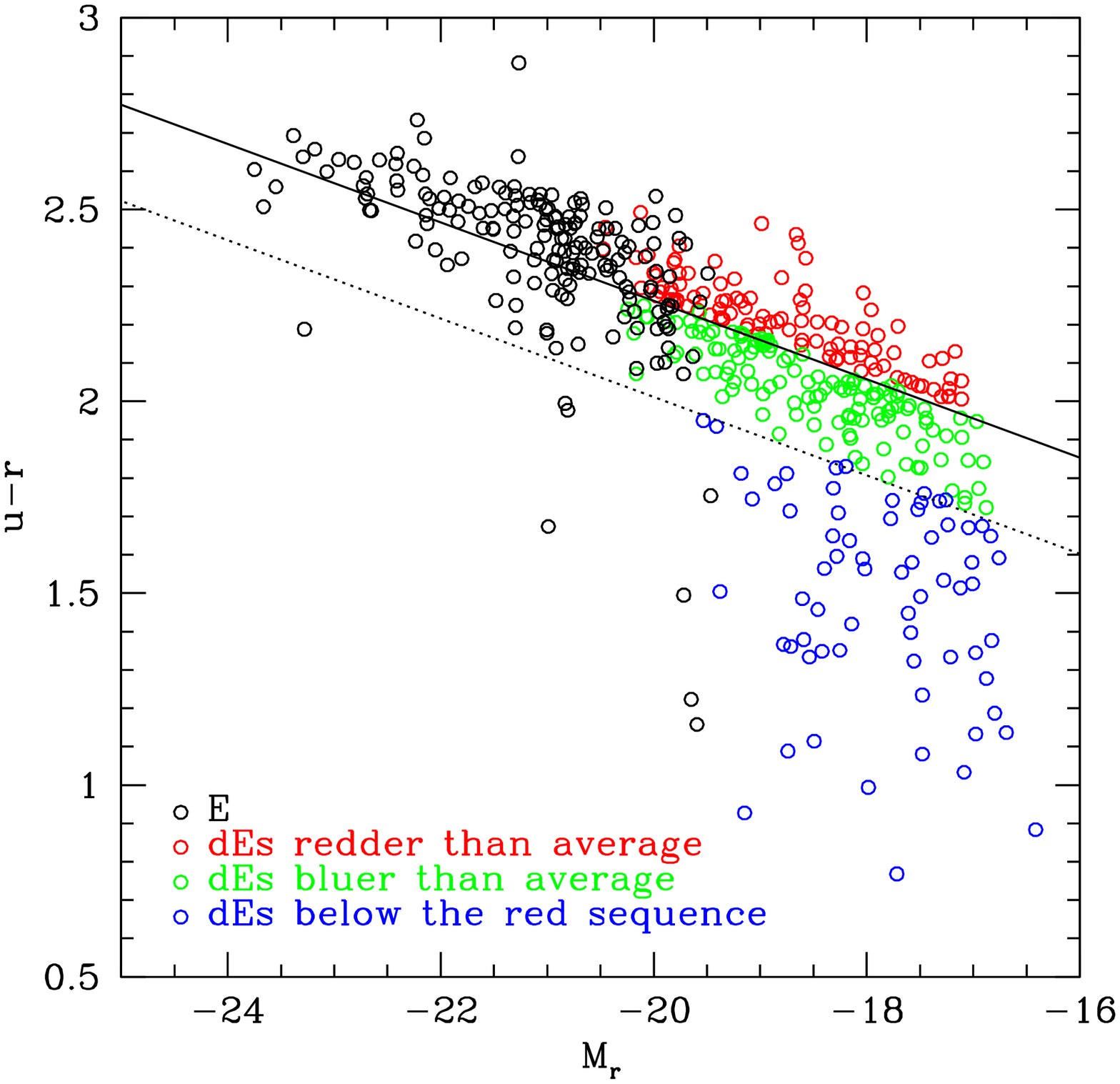}
\caption{The CMD in SDSS $u$-- and $r$--bands. The solid line is a fit
  to the red sequence of dEs in the Virgo cluster from
  \cite{lis08}. The dotted line is the same, but shifted by 0.25 mag
  to the blue. These lines are used to define three subsamples of dEs.
  The color has a typical error in the range $0.10-0.15$ mag.}
\label{cmd_sdss}
\end{figure}

Figure \ref{cmd} has shown that Es and a majority of dEs form a red
sequence in a CMD. Now we want to compare this red sequence to the one
found for dEs in the Virgo cluster and use the corresponding fit to
further subdivide our dE sample. In Fig. \ref{cmd_sdss} we plot a CMD
in SDSS $u$-- and $r$--bands (we use the SDSS filters here, since the
Virgo study has been performed with SDSS data). The solid line is the
fit to the red sequence of dEs in the Virgo cluster obtained in the
study of \cite{lis08}. We use this fit rather than a fit to our own
data, because it is based on a dE sample covering a much larger range
in magnitudes. The dEs in our sample are in good agreement with the
Virgo fit, while the majority of Es are shifted towards redder
colors. A nonlinear color-magnitude relation for a combined sample
of giant and dwarf early-type galaxies has been reported by \cite{jan09},
while linear color-magnitude relations have
been found in recent photometric studies of nearby galaxy
clusters \citep{mis08,mis09,sca08}. We use this fit to further
subdivide our dE sample, which is indicated in Fig. \ref{cmd_sdss} by
the red, green, and blue circles. We define three subsamples based on
the fit and on a parallel shift of the fit by 0.25 mag to the blue
(dotted line): dEs with redder colors than expected based on their
magnitude ('red dEs`, 103 objects), the subsample of dEs with bluer
colors ('intermediate dEs`, 126 objects), and the subsample of dEs,
which exhibit rather large separation from the red sequence as
discussed above ('blue dEs`, 66 objects).

This approach is based on the assumption that the interaction of the
galaxies with the cluster causes them to move from the blue cloud to
the red sequence on the CMD. This process takes some time and the
change in color of a galaxy is not occurring at once. In this picture,
the redder dEs are subjected to this process for a longer time than
the bluer dEs and are also believed to have an older stellar
population (see the discussion of corresponding results in Sect.
\ref{dist}). By separating the dEs along the red sequence fit, we
are trying to investigate this process in the following.

\section{The distribution of dEs in A901/902}\label{dist}
\begin{figure*}
\centering
\subfigure
{
    \includegraphics[width=9cm]{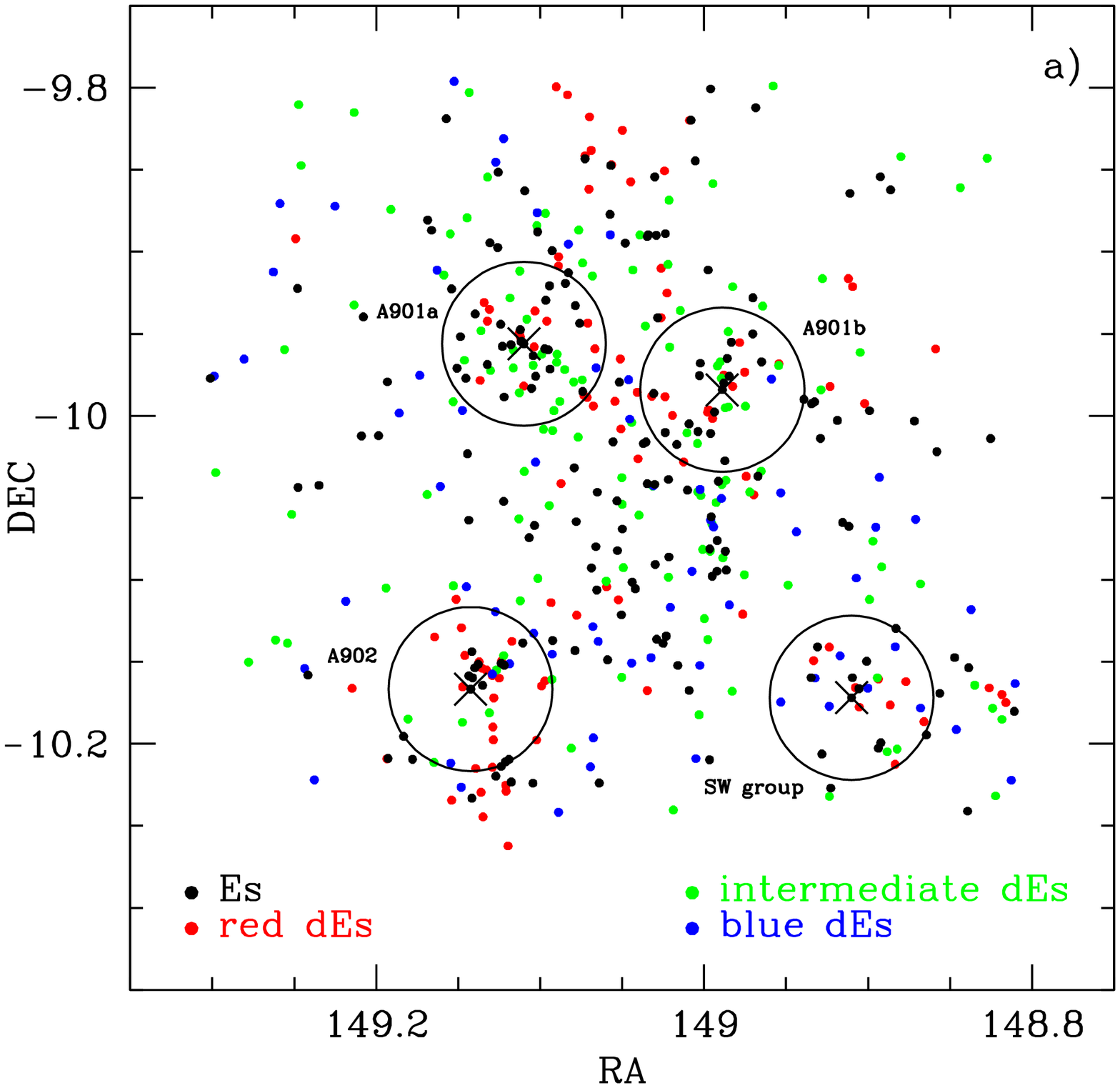}
}
\hspace{-0.5cm}
\subfigure
{
    \includegraphics[width=9cm]{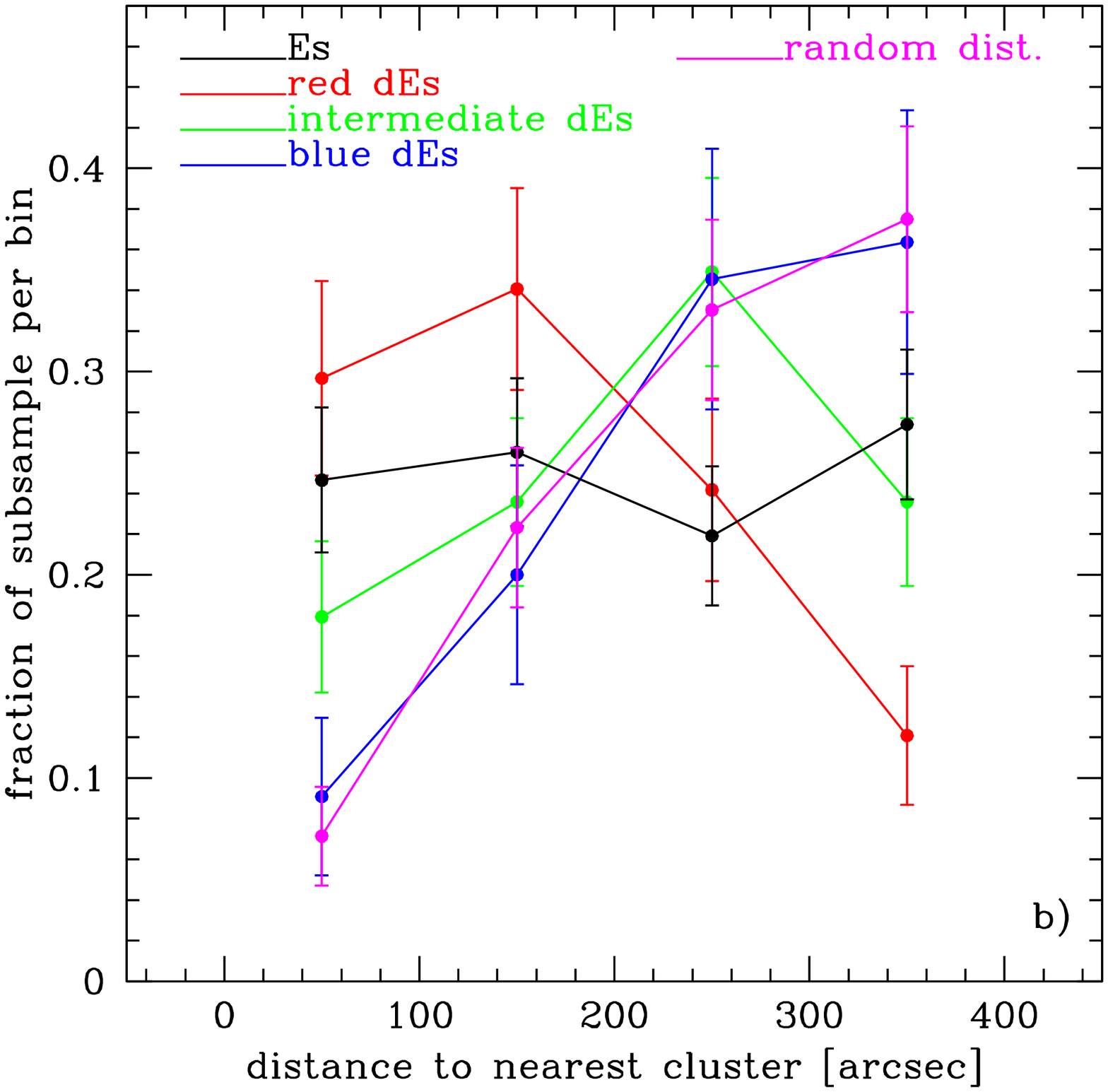}
}
\caption{{\it a)} A map of the multiple-cluster system A901/902
  showing the distribution of all early-type galaxies. The different
  colors represent the subsamples as defined in
  Fig. \ref{cmd_sdss}. The circles have a radius of $\sim500$ kpc. The
  centers of the four subcomponents of the multiple-cluster system are
  indicated by black crosses. {\it b)} The fraction of each subsample
  and a random distribution of galaxies in four distance (to the
  nearest cluster) bins.}
\label{maps}
\end{figure*}

The distribution of early-type galaxies in the multiple-cluster system
is shown in Fig. \ref{maps}. Four distinct subcomponents have been
identified in A901/902 \citep[A901a, A901b, A902, and the South West
  group (SW group),][]{gra02,hey08}, which are indicated by black
crosses in Fig. \ref{maps}a. The circles have a radius of $\sim500$
kpc. The galaxies are clearly concentrated towards the four structure
centers. In particular the red dEs avoid the region between the
subcomponents (central region in Fig. \ref{maps}a). In a next step we
determine the projected distances (in arcsec) of all galaxies to the
nearest subcomponent. We bin these distances in four distance ranges
and show the corresponding fraction of each subsample within these
bins in Fig. \ref{maps}b. Also shown is the result for a random
distribution of galaxies. It is based on the average of 1000 random
distributions of 180 galaxies in the cluster system. As indicated by
Fig. \ref{maps}a, the red dEs are strongly concentrated towards the
centers of the subclumps, with a fraction of $\sim56\%$ within the
first two bins (corresponding to $\sim560$ kpc). Only $\sim35\%$ of
the intermediate dE population is located within the same radius.
A KS--test shows that the probability that the red and
intermediate subsamples stem from the same parent distribution is
$<0.5\%$. The distribution for Es is rather flat, i.e., still
strongly concentrated towards the centers as opposed to a random
distribution (magenta line), but to a lesser degree than red dEs.
The population of blue dEs shows no concentration towards the centers
and their distribution is the same (within errors) as the one expected
for a unrelated field population (magenta line).

\begin{figure}
\centering
\includegraphics[width=9cm]{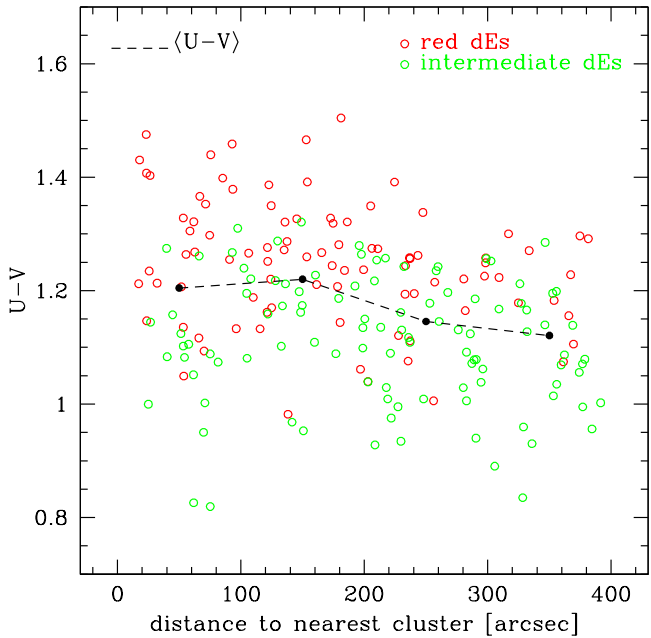}
\caption{The total $U-V$ color of red and intermediate dEs as a
  function of distance to the nearest cluster center. The filled
  circles show the average colors in four distance bins.}
\label{cadis}
\end{figure}

The difference between the distributions of red and intermediate dEs
corresponds to a color-density or color-radius relation, after the
correlation between colour and luminosity has been removed. Hence,
the scatter of the red sequence for dEs in Fig. \ref{cmd_sdss} is
related to the distribution of the galaxies in the cluster region. Figure
\ref{cadis} shows the $U-V$-color as a function of distance to the
nearest cluster center for red and intermediate dEs. The overall
color-density relation of the dEs on the red sequence is in fact
quite weak, showing the importance of the normalization by luminosity.

The fact that early-type dwarfs are strongly concentrated towards the
centers of their host clusters was also found in studies of local
galaxy clusters \citep{bin87,fer89,lis07}. In particular, nucleated
dEs exhibit a stronger concentration towards the cluster center than
the typically fainter non-nucleated dEs. A finding, we cannot verify
with our data (see below). A significant trend of dE color with
clustercentric radius in the Coma cluster, where redder dEs tend to be
closer to the cluster center, was reported by \cite{sec96}; although
the scatter of the data points was rather large. A similar result
based on the color-magnitude relations for different dEs
subpopulations was also found by \cite{lis08} for the Virgo cluster.
Moreover, \cite{wol07} find evidence for a relation between the
residuals from the color-magnitude relation and local galaxy density
for objects in the range $-18<M_V<-20$ mag using the COMBO--17 data
for A901/902. On the other hand, \cite{san08} did not find a color
gradient among red dwarf galaxies in a large compilation of galaxy
clusters mapped by the SDSS. In general the gradients found are weak
and only significant taking into account projection effects and
assuming that these will always weaken an intrinsic color gradient.
Finally, age gradients in the sense that older dEs are closer to the
cluster centers were found in the Virgo \citep{mic08} and Coma clusters
\citep{smi09}. These studies also show that the dEs lying closer
to the cluster centers tend to be more metal poor. Together with the
results presented in our study, this indicates that the color
differences among dEs with the same luminosity are more related to
their ages than to their metallicities.

\section{The compactness of dEs}\label{compa}
Many dEs are found to harbor nuclei, which most likely are massive
compact star clusters like the nuclei in nearby low-luminosity spirals
\citep{kor93,phi96,mat99,boe02}. The nuclei in dEs are located at or
near the center of the luminosity distribution of the host galaxy
\citep{bin00,bar03} and have sizes in the range $\la2\sim60$ pc
\citep{geh02,cot06}. In models of dE formation, nucleated dEs (or
dENs) are predicted to be more strongly concentrated towards the
cluster centers than non-nucleated dEs, since the former are more
robust against the impact of the environment \citep{moo98}. This
property of the dEN population was reportedly found in the Virgo
cluster \citep{van86,bin87,lis07}, but the result is still
controversial \citep{cot06}.

\begin{figure*}
\centering
\includegraphics[width=9cm]{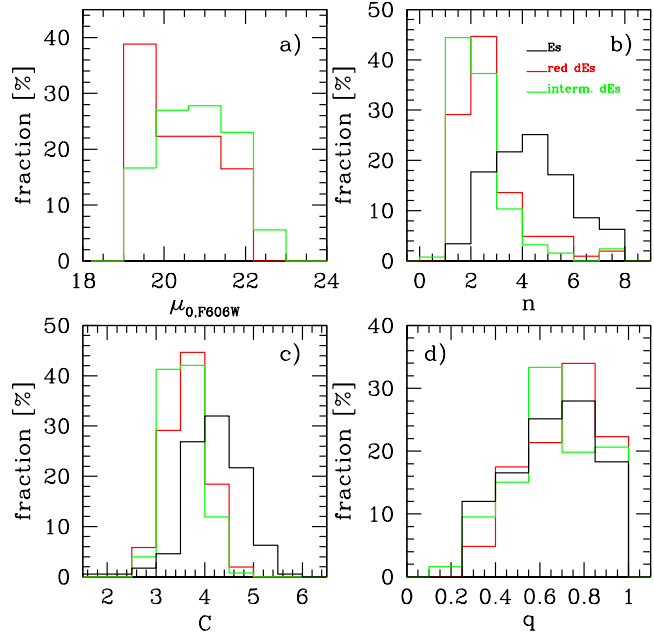}
\caption{The distributions of different measures of compactness and
  roundness for red dEs (red histograms), intermediate dEs (green
  histograms), and giant Es (panels $b-d$, black histograms). {\it a)}
  Central surface brightness, $\mu_{\rm 0,F606W}$, as defined in
  Sect. \ref{defsam}; {\it b)} S\'ersic index $n$; {\it c)}
  concentration index $C$; {\it d)} axis ratio $q$, defined as
  semi-major axis divided by semi-minor axis.}
\label{compp}
\end{figure*}

Due to their small sizes, the nuclei in our dEs cannot be resolved and
their identification is therefore very uncertain. However, in order to
investigate, whether the more robust dEs exhibit a stronger
clustercentric concentration, we can rely on other measures of
compactness, indicating the possible presence of a nucleus. We have
shown in Fig. \ref{maps}b that the red dEs are more centrally
concentrated within the cluster than the intermediate dEs. Now we
test, whether the former are more compact than the latter. In
Fig. \ref{compp} we show the histograms of different compactness
indicators for our subsamples. Figure \ref{compp}a shows the
distributions of the central surface brightness for red and
intermediate dEs. The two distributions are indeed different: red dEs
have higher central surface brightnesses with a median of
$\langle\mu_{\rm 0,F606W}\rangle=20.27$ mag/arcsec$^2$ compared to
$\langle\mu_{\rm 0,F606W}\rangle=20.86$ mag/arcsec$^2$ for intermediate dEs. A
KS-test shows that the probability that the two samples are drawn from
the same parent distribution is $<1\%$. For the S\'ersic index $n$,
shown in Fig. \ref{compp}b, we obtain a similar results: the median
values are $\langle n\rangle=2.45$ and $\langle n\rangle=2.12$ for red
and intermediate dEs, respectively. The KS probability is
$<0.1\%$. The result is less pronounced for the concentration index C
(Fig. \ref{compp}c), derived using the CAS code
\citep{con00,hei09}. For C we find $\langle C\rangle=3.61$ for red dEs
and $\langle C\rangle=3.55$ for intermediate dEs, with a KS
probability of $\sim3\%$. These findings indicate that dEs located
close to the cluster centers are on average redder and more compact
than their counterparts at larger radii. Moreover, the results also
support the expectation that nucleated dEs should be more centrally
concentrated within the cluster than non-nucleated dEs. This view is
also based on the fact that our two dE subsamples show significant
differences in terms of $\mu_{\rm 0,F606W}$ and $n$, which are better
tracers of the presence of a nucleus than $C$, since they are more
sensitive to the very central part of the surface brightness profiles
of the objects. Finally, Fig. \ref{compp}d shows the distributions of
the axis ratios ($q$). Red dEs tend to be rounder
($\langle q\rangle=0.72$) than intermediate dEs ($\langle q\rangle=0.64$,
KS probability $\sim7\%$). In fact, dENs are found to be rounder on
average than non-nucleated dEs \citep{bin95}, which provides further
support for our finding that more compact (and possibly nucleated) dEs
exhibit a stronger concentration towards the cluster centers than the dE
population in general.

\section{Discussion}\label{discu}
The main finding of our study is a color-density relation among dEs
in the sense that for galaxies with the same luminosity the redder
objects are located closer to the cluster center than the bluer
ones. In addition, the redder dEs tend to be more compact and
rounder. These findings indicate that fundamental properties of dEs
are affected by the interaction with the cluster environment. In
particular the star formation history (color-density relation) and
the structure (compactness and roundness) of dEs seem to depend on
their specific location within the cluster region.

Before discussing possible processes that could have led to our
findings, we have to point out that the dE population in clusters
covers a wide range of luminosities, reaching magnitudes as faint as
$M_B\sim-10$ mag \citep{sab03}. It is therefore likely that different
formation processes are at work and that the faint dEs have a
different origin than their brighter counterparts. Since our sample
is dominated by rather luminous objects ($M_B\leq-16$ mag), our
discussion concentrates on the possible processes leading to the
formation of $bright$ dEs.

\subsection{Ram pressure stripping}
Ram pressure stripping describes the effect of the inter-cluster
medium (ICM) on the inter-stellar material (ISM) of a galaxy moving
through the cluster \citep{gun72}. If the density of the ICM is
high enough, its pressure can lead to the total or partial removal
of the ISM from the moving galaxy. Ram pressure stripping is more
efficient in lower mass galaxies and has therefore been invoked as
a possible process that terminates star formation in dwarf irregulars
and turns them into quiescent dwarf ellipticals in clusters
\citep{lin83,aba99,zee04}. Observational evidence for ram pressure
stripped galaxies in nearby clusters has been presented in recent
studies \citep[e.g.,][]{vol04,kan08}. Since the density of the ICM is
increasing towards the cluster centers, it is plausible that galaxies
with smaller clustercentric distances experience a stronger impact
from ram pressure stripping than galaxies at larger radii. This can
affect their star formation histories. The ISM of galaxies
close to the cluster centers might be removed faster and more
efficiently, and as a consequence their star formation terminates
earlier compared to their counterparts at larger radii. The former
would then be dominated by an older stellar population than the latter.
Thus, our finding of a color-density relation is consistent with this
picture. The fact that we find a stronger relation using color
residuals, i.e. taking into account the luminosity of the galaxies,
could be due to the fact that the strength of ram pressure stripping
also depends on the mass of the affected galaxy. On the other hand, our
additional results that the compactness and axis ratios of dEs increase
towards the cluster centers are more difficult to reconcile with this
scenario. It is not clear how the mere removal of gas can cause these
structural signatures. For a discussion of possible explanations and
related issues see \cite{bos08a}.

\subsection{Galaxy harassment}
Galaxy harassment describes the total morphological transformation of
disk galaxies into quiescent spheroidals in a galaxy cluster through
repeated encounters with more massive galaxies and the cluster's tidal
field \citep{moo98}. Hence, this process is believed to be able to
transform late-type disk galaxies into dEs. Further N-body simulations
have shown \citep{mas05} that in some cases the transformation
remnants can maintain parts of their original disk structure within a
spheroidal component. Such hidden disk structures have been found in a
number of bright dEs in different nearby clusters where spiral or bar
features have been observed \citep{jer00,bar02,gra03b,der03,lis06}.
The model can also account for the presence of compact nuclei in the
centers of many cluster dEs \citep{san84,fer94}, which have also been
observed in a large fraction of nearby late-type disk galaxies
\citep{car97,boe02}. Moreover, the galaxy harassment model predicts that
the surviving ensemble of dEs exhibit a color- or age- (also metallicity)
density relation within the galaxy cluster and that the fraction of
nucleated dEs increases towards the cluster center \citep{moo98}. The
color-density relation found for the A901/902 dEs is therefore
consistent with the harassment scenario. Our finding that the redder dEs
are on average more compact and rounder than bluer dEs also agrees with
the model predictions, since nucleated dEs have been found to be rounder
than non-nucleated dEs \citep{bin95}. Therefore the model predicts an
ellipticity gradient among dEs, which can be inferred from our results.
On the other hand, we do not find a gradient in terms of effective radius,
which is also predicited by the harassment model \citep{mas05}. The
remnants of the interactions in the outskirts of the cluster are found to
have larger effective radii by a factor of $\sim2$ compared to the ones
at smaller clustercentric distances. We do not find a corresponding
difference between red and intermediate dEs, which have roughly the same
average effective radii. Finally, we want to point out that support for
the assumption that late-type, intermediate-mass disk galaxies are affected
by the cluster is provided by a study of the environmental dependence of
the stellar mass-size relation in A901/902 using STAGES data \citep{mal09}.
The results indicate that almost all galaxies show no difference in the
stellar mass-size relation between cluster and field. However, the only
galaxies to show a possible difference were the class of intermediate-mass
($\log M<10$ M$_{\odot}$) spirals, where the spirals in the cluster were
found to have a smaller mean effective radius than their counterparts in
the field at the $2\sigma$ level.

\subsection{Preprocessing in galaxy groups}
Galaxies are not necessarily joining a cluster as individuals, but
as part of a galaxy group. Hence, part of their evolution might
take place in the group environment prior to their accretion to a
galaxy cluster. Therefore, these galaxies were {\it preprocessed}
before they experienced the impact of the interaction with the
cluster environment \citep{cor06,kau08,lii09}. It has been shown by
\cite{may01b} that star forming dwarf galaxies can be transformed
into early-type dwarfs in a Milky Way sized halo by tidal stirring.
These models also predict relations between galaxy properties and
their distance to the parent galaxy \citep{may01a}. The Abell
901/902 cluster system is composed of four distinct subcomponents,
whose centers are indicated in Fig. \ref{maps}a and were used to
define the radial distributions in Fig. \ref{maps}b. However, only
the SW group can be regarded as a typical galaxy group, the other
substructures are more massive. Their central velocity
dispersions are of the order of $\sigma\sim600-900$ kms$^{-1}$
\citep{hei09}, which place them between the Virgo cluster
\citep[$\sigma\sim400-750$ kms$^{-1}$,][]{bin87} and the Coma
cluster \citep[$\sigma\sim900$ kms$^{-1}$,][]{the86}. The
SW group is smaller ($\sigma\sim500$ kms${^-1}$) and contributes
13 ($\sim13\%$) and 18 objects ($\sim14\%$) to our red and
intermediate subsamples, respectively. Their distribution with
respect to the center of the SW group is very similar to the one
shown for the entire subsamples in Fig. \ref{maps}b. The SW group
might therefore represent an example of a (rather massive) galaxy
group delivering preprocessed dwarf galaxies to a larger galaxy
system. However, we want to point out that the remnants in
simulations of group related processes are commonly rather small,
low mass dwarfs comparable to the dwarf spheroidal galaxies in the
Local Group and therefore very different from the bright dEs in
our sample.\\

The fact that dEs are very abundant in galaxy clusters, which can
strongly affect their members, does not necessarily prove that they
are the results of transformation processes. It is quite possible
that some dEs belong to the primordial cluster population and
experienced the impact from the cluster environment throughout their
evolution leading to the objects we observe today. On the other hand,
it is clear that galaxies can be profoundly affected by the
interactions with a galaxy cluster. The agreement between our findings
and the predictions of the cluster related processes discussed above
is consistent with a scenario in which a considerable fraction of dEs
in A901/902 have been affected by their environment.

\section{Summary and conclusions}\label{sum}
We have used data from the multiwavelength survey STAGES \citep[Space
  Telescope A901/902 Galaxy Evolution Survey,][]{gra09} to study the
population of bright dEs in the multiple-cluster system Abell
A901/902. Our definition of dEs is based on the identification of
early-type galaxies through visual classifications and a
separation of dwarfs and giants using a cut in central surface
brightness at $\mu_{\rm 0,F606W}=19$ mag. We subdivide our dE sample into
red, intermediate, and blue dEs according to their position on a CMD
using $M_r$ and $u-r$. The corresponding dE subsamples exhibit very
different distributions within the cluster region with respect to the
distance to the nearest cluster center. The red dEs are strongly
concentrated towards the cluster centers, even stronger than giant Es,
while the distribution of the intermediate dEs is comparable to the
one of Es and the blue dEs avoid the cluster centers. The distribution
of the latter sample is similar to the one of a field population and
we cannot rule out that a large fraction of objects in this sample are
contaminants seen in projection within the cluster region. The
different distributions of red and intermediate dEs is interpreted as
a color-density relation with respect to the cluster centers in the sense that
for dEs with similar luminosities the redder galaxies are more
centrally concentrated in the cluster than the bluer
objects. Furthermore, we find evidence that the red dEs are more
compact, indicating the presence of a nucleus, and rounder than the
bluer subsample of dEs.

The relation between fundamental properties of dEs and their location
with respect to the cluster centers suggests that the formation of
dEs is strongly affected by cluster related processes. We
therefore compare our results to the model predictions based on
processes that are believed to take place in a cluster environment.
Our findings are consistent with a number of these predictions and
indicate that a considerable fraction of dEs in A901/902 are the
result of transformations. It is plausible that several processes are
involved and affect the evolution of the cluster galaxies to some
degree. However, our data are not sufficient to determine the
contributions of the different factors shaping the dE population in
A901/902 and to provide evidence that transformations are involved at
all. On the other hand, we find good agreement between our findings
and the predictions of the galaxy harassment model indicating that
processes described in this scenario are most likely contributing to
the formation and evolution of at least some dEs in A901/902.

\begin{acknowledgements}
We thank the anonymous referee for a helpful and constructive report.
CW and MEG were supported by an STFC Advanced Fellowship.
SJ, AH, and IM gratefully acknowledge support
from NSF grant AST--0607748, LTSA grant NAG5--13063, as well as
programs HST GO--10395, HST GO--10861, and HST GO--11082, which were
supported by NASA through a grant from the Space Telescope Science
Institute, which is operated by the Association of Universities for
Research in Astronomy, Incorporated, under NASA contract NAS5--26555.
MB was supported by FWF grant P18416. Support for STAGES was provided
by NASA through GO--10395 from STScI operated by AURA under NAS5--26555.
\end{acknowledgements}

\end{document}